\documentclass[12pt]{article}
\def\lsim{\mathrel{\rlap{\lower4pt\hbox{\hskip1pt$\sim$}}
    \raise1pt\hbox{$<$}}}         %less than or approx. symbol
\def\gsim{\mathrel{\rlap{\lower4pt\hbox{\hskip1pt$\sim$}}
    \raise1pt\hbox{$>$}}}         %greater than or approx. symbol

\title{Nuclear Physics Neutrino PreTown Meeting: \\
Summary and Recommendations}
\author{}
\date{\vskip -2 in}

\begin{document}
\maketitle

\section{Introduction}

In preparation for the nuclear physics Long Range Plan exercise,
a group of 104 neutrino physicists met in Seattle September 21-23
to discuss both the present state of the field and the  
new opportunities of the next decade.  This group included a substantial
fraction of the US nuclear physics neutrino community.  The meeting was organized
around five working groups: solar neutrinos; supernova neutrinos
and the supernova mechanism; reactor and accelerator neutrinos;
underground laboratories (including atmospheric and high energy
neutrinos); and neutrino mass (double beta decay and tritium beta
decay).  Plenary speakers summarized the status of field, while
the important question of future directions was tackled by the
working groups.  In the third day of the meeting the working
groups presented their conclusions in a plenary session that 
concluded with a group discussion of future opportunities and
priorities. \\

\noindent
While discussions were wide ranging, three broad themes surfaced
frequently: \\

\noindent
$\bullet$ Although nuclear physics has a long tradition in neutrino
physics -- landmark events include introduction of the neutrino to
conserve energy in $\beta$ decay, the precision $\beta$ decay tests
that contributed to the experimental foundations of the standard
model, and the chlorine solar neutrino experiment, which started 
the field of neutrino astrophysics -- now is a time of special
opportunity.  The neutrino physics done by nuclear physicists 
lies at the intersection of two major intellectual revolutions.
One of these is the nature of physics beyond the standard model.
Measurements of the atmospheric neutrino flux have provided the
first clear evidence that the standard model is incomplete.
The most naive interpretation of the neutrino mass difference derived
from these measurements suggests a seesaw mass of about $10^{15}$
GeV, remarkably close to the grand unification scale.  With SNO
and other new solar neutrino experiments, a new generation of 
massive double beta decay experiments, and opportunities like
KamLAND, MiniBooNE, and K2K to probe oscillations with
accelerator and reactor neutrinos, the
next decade should be a time of rich discovery.  The 
pattern of neutrino masses and mixings should be revealed, thereby
guiding the efforts of theorists to formulate a new standard model.\\
\hspace*{1cm}A second revolution is occurring in astronomy and astrophysics:
rapid technological advances are allowing observers to probe
the universe with increasing precision and in a  
breadth of wavelengths.  Among the problems that may be clarified
in the next decade are the nature of dark matter and dark energy;
the origin of the elements; the evolution of structure;
and the physics of extreme environments, including supernova 
explosions, neutron stars, and cosmic rays of energy in excess of
$10^9$ TeV.  Neutrino physics is central to much of 
this physics.  The atmospheric neutrino results already demand 
that neutrinos are as important as the stars in their contribution
to the mass of the universe.  Neutrino masses will become important
cosmological parameters as more precise microwave and large-scale
structure maps of the universe are made.  Neutrinos control the proton-neutron
chemistry of supernova ejecta in which we believe the heavy
r-process elements are synthesized, and are directly involved in
the synthesis of certain nuclei.  Just as solar neutrinos allow us
to probe conditions in the solar core, the supernova neutrino
``light curve'' may help us better understand such stellar explosions,
and may carry information on the state of the high-density nuclear matter in
the protoneutron star.  High energy neutrinos may
allow us to look inside the universe's most energetic central
engines.  Conversely, these new environments will 
extend our tests of basic neutrino physics: 
matter effects for supernovae neutrinos can dramatically enhance
oscillations that would otherwise be unobservable, for example.\\
\hspace*{1cm}Nuclear physics has an opportunity to play major roles in these
two revolutions. \\

\noindent
$\bullet$ In the last decade decisions to expand the boundaries of
nuclear physics -- to the substructure of the nucleon (JLab) and 
to the phases of nuclear matter at extremes of temperature and
density (RHIC) -- helped to revitalize the field.  It is important
to maintain this momentum by exploring new frontiers in the next
decade.  A new initiative in nuclear astrophysics, which would 
encompass the neutrino physics discussed above, is an outstanding
opportunity for the field.  Such an initiative could include
other physics now under discussion in the Long Range Plan
process, such as the Rare Isotope Accelerator, with its strong
program of laboratory astrophysics.  Clearly the impact of future
studies of the nuclear structure of the r-process will be 
enhanced if nuclear physicists can simultaneously understand
the supernova neutrino physics that controls the r-process
path.  Progress on both fronts would hopefully lead to a standard
model of the core-collapse supernovae and the associated 
nucleosynthesis.  The analogy with the solar neutrino problem,
where the importance of laboratory measurements of $pp$ chain
reaction rates was enhanced by efforts on solar neutrino 
detection and on the standard solar model, is quite striking. \\

\noindent
$\bullet$ Despite the interest in and promise of nonaccelerator
neutrino physics, US physicists have had to overcome obstacles in mounting
major experiments.  Important ideas have come
from the US community (e.g., SAGE/GALLEX and SNO), but the
detectors have been built elsewhere, with US scientists as
participants.  Similarly, though US physicists are active in the
field, no double beta decay experiment 
is currently running in the US.  The reasons for this situation
are complex.  In the case of SNO, the Canadians owned the heavy
water.  In the case of SAGE/GALLEX, community recommendations to
fund these experiments were not followed.  A factor in siting 
some double beta decay experiments overseas was the quality
of available underground sites.  Despite this history, there are recent
signs that support for the field is increasing: significant
agency investments have been made in SNO, Borexino, SAGE/GALLEX,
and KamLAND. \\
\hspace*{1cm}Europe and Japan have moved ahead of the US in important respects.
Italy's Gran Sasso laboratory was created to foster underground
experiments in Europe.  It has become a major center,  
encouraging new ideas in underground physics and drawing
experiments from across Europe and elsewhere.  It is currently
oversubscribed, prompting discussions of expansion.  In Japan
the Kamioka proton decay experiment, contemporaneous with 
the US IMB detector, was followed by SuperKamiokande, an 
effort that has had a profound influence due to its solar and 
atmospheric neutrino discoveries.  There was no US followup
to IMB, the first experiment to uncover an anomaly in the
atmospheric neutrino flux. \\

The recent discoveries in neutrino physics provide the US nuclear
physics community with an opportunity to rethink its strategy in 
neutrino physics and related areas of nuclear astrophysics.  
The benefits to nuclear physics are not confined to the physics results:
The field is very popular with students and thus is an 
important aid to recruitment.  The ultimate success of any field
depends on the quality of the students it attracts.  Physics is
also increasingly interconnected, with discoveries in one area
affecting progress in others.  Neutrino physics is an outstanding
example, with relevance to astrophysics, cosmology, and
particle physics.  Results from experiments like SNO have the
potential to influence all of physics, and thus to contribute to
the stature of our field.

\section{Principal Recommendations}

While there are additional recommendations discussed in the working
group summaries, the following four could form the basis of a 
neutrino initiative for the next decade: \\ 

The current generation of solar neutrino experiments is expected
to produce exciting results, including the first separation of
the $^8$B neutrino flux into electron and heavy flavors.
The next major push in this field
must involve active detectors capable of determining the flux 
and flavor of the low-energy $pp$ and $^7$Be neutrinos.  Most candidate 
solutions to the solar neutrino puzzle affect this portion of the
spectrum in distinctive ways.  Several detectors with
the necessary charactistics are well along in development.
Remarkable progress in
neutrinoless double beta decay over the past two decades -- a
factor of two increase in lifetime limits every two years -- 
has now reached a fundamental limit, $\sim 10^{25}$ years, imposed
by current detector sizes ($\sim$ 10 kg).  There are now urgent reasons for 
probing Majorana neutrino masses at the 0.03-0.10 eV level, 
requiring ton masses of the parent nucleus.  Several excellent 
experiments have been proposed, some of which are technically
well developed.  Some of these exploit idle Russian
enrichment capabilities.
  
One obstacle to increasingly precise neutrino experiments is the absence
of a US deep underground laboratory, isolated from cosmic rays
and other background sources.  This has forced
US experimentalists to mount their experiments elsewhere,
or to manage in less than optimal surroundings, such as active
mines.  The announced closing of the Homestake Mine in 16 months,
coupled with the interest of the state of South Dakota in 
converting this to a national scientific facility, could provide
a very deep (4850-8000 ft) hard-rock site.  San Jacinto
remains the best studied possibility for creating a site with
horizontal access.  The Sudbury Neutrino Observatory could be
further developed as a multipurpose underground laboratory
for North America. \\

\noindent
{\it Recommendation \#1:  Neutrino experiments in nuclear physics
are making fundamental contributions to our understanding of
the mass, mixing, and charge conjugation properties of neutrinos.
The new discoveries are crucial to our field, affecting our 
understanding of nucleosynthesis, supernovae, and many other
phenomena, and to astrophysics, particle physics, and cosmology. \\
  
\noindent
$\bullet$ Nuclear physics must build on its low energy neutrino
successes, fully exploiting the existing detectors,
while, at the same time, preparing to undertake the
next generation of solar neutrino/supernova neutrino and double
beta decay experiments.  Several interesting next-generation 
experiments have been proposed.  It is imperative to move these
projects quickly through the R\&D phase, so that the most 
promising detectors can be identified and launched as 
full-scale experiments.  The community and agencies should work
together to accomplish this goal. \\ 

\noindent
$\bullet$ To satisfy the background requirements of new
solar/supernova neutrino and double beta decay experiments,
the nuclear physics community should spearhead an effort
to create a deep underground multipurpose laboratory.
Because this national facility could also serve the needs of dark matter
and nucleon decay experiments, it is important to involve colleagues
from particle and astrophysics.  The urgency of one of the proposals
(Homestake) requires that the community move now to define the
merits and attributes of such a facility.} \\

The Spallation Neutron Source at Oak Ridge National Laboratory is 
now under construction.  One byproduct of this facility is 
a stopped pion neutrino source of unusual characteristics: 
a flux in excess of that achieved at LAMPF, a pulsed time structure
similar to that of the ISIS facility at Rutherford Laboratory,
and an unusually low contamination of $\bar{\nu}_e$ (important
for oscillation tests).  The ORLAND collaboration has proposed 
exploiting this $\sim$ \$1B new accelerator by constructing a neutrino 
bunker near the beam stop.  A variety of neutrino experiments --
new oscillation tests, measurements of neutrino-nucleus cross
sections important to supernova physics, tests of isoscalar axial
currents -- could be tackled with such a facility. \\

\noindent
{\it Recommendation \#2:  Nuclear physics should construct a
neutrino bunker at the SNS beam stop so that the pulsed neutrino
flux can be exploited in future experiments.  It is important 
that the bunker be situated as close as possible to the beam
stop.  Such a facility would allow community collaborations to 
propose detectors and conduct experiments.} \\
  
A major nuclear astrophysics challenge is to construct and experimentally
verify a standard model of core-collapse supernovae.  This problem 
encompasses nuclear theory, astrophysics, and computer
science, requiring modeling of the
nuclear equation of state up to at least four times nuclear density;
of hydrodynamics, convection, and shock wave propagation;
and of the neutrino-nucleus microphysics that we
believe is crucial to both the explosion mechanism and associated
nucleosynthesis.  Supernova physics has already been an important 
stimulus to nuclear structure, motivating a great deal of recent
shell model work (for example, finite temperature shell model
Monte Carlo studies of Gamow-Teller strength distributions and
level densities).  A standard supernova model 
is needed to make full use of the
the nuclear structure along the r-process path that RIA will provide.
It is also needed to understand detailed abundances patterns that
have recently come from Hubble Space Telescope and other studies of metal-poor
stars enriched in r-process metals.  Finally, it is essential 
if we are to exploit the next galactic supernova as a laboratory for
new neutrino physics: unique kinematic neutrino mass and matter-enhanced
oscillation tests are possible with supernova neutrinos. \\

\noindent
{\it Recommendation \#3:  The supernova mechanism is an outstanding
computational and theoretical ``grand challenge'' problem in nuclear
physics and astrophysics.  A new theory initiative should be launched to make 
progress toward a multi-dimensional model with realistic neutrino transport
and nuclear microphysics.  An important component of this effort 
is improvements in nuclear structure methods for neutrino-nucleus
cross sections.} \\

Neutrino physics lies at the intersection of nuclear, particle,
and astrophysics.  Collaborations often cross subfield lines:
the different experimental skills within these subfields are a source of
vitality for neutrino physics.  One example is the AMANDA detector 
(and its proposed successor ICECUBE) for observing high-energy astrophysical
neutrinos.  AMANDA, which exploits the extensive experience in high
energy physics with water Cerenkov detectors, will use neutrinos to probe
active galactic nuclei and other astro-particle accelerators,
very much as nuclear physicists are using solar neutrinos to probe
the solar core.  A second example is the next-generation
proton decay detector being planned by particle physics: 
nuclear physics interest might
focus on exploiting such a massive detector for solar or 
supernova neutrino detection.  A third example is the proposed neutrino
factory.  At the time of the next long range plan nuclear physics
will need to consider the opportunities for nuclear structure-function
measurements this facility might provide.  It is important for
nuclear physics to have some involvement in these and other similar
projects because they are relevant technologically and intellectually
to nuclear physics. \\

\noindent
{\it Recommendation \#4:  Nuclear physics should continue to 
support members of our community who collaborate on relevant 
neutrino experiments funded primarily by other subfields.}
  
\section{Working Group Summaries}

The five working groups addressed many of the questions that have
been posed to guide the Long Range Plan process.  Here
the responses are summarized.

\subsection{Solar Neutrino Working Group}
\noindent
{\it What scientific questions is this subfield trying to answer?} \\  
There are five principal questions driving the field:\\
$\bullet$ What nuclear physics governs energy production in our 
sun's core and in other stars?  Is our understanding of stellar 
evolution quantitative? \\
$\bullet$ What is the origin of the solar neutrino problem? \\
$\bullet$ Do electron neutrinos oscillate and, if so, to what? \\
$\bullet$ What constraints on neutrino masses and mixing angles
can be extracted from solar neutrino experiments? \\
$\bullet$ Do neutrinos have other non-standard-model properties,
such as magnetic moments or flavor-changing interactions?

The pattern of solar neutrino fluxes that has emerged from current experiments,
combined with the atmospheric neutrino evidence for neutrino mass,
strongly suggests that the solar neutrino problem is due
to neutrino oscillations.  The current interpretation of the
SuperKamiokande atmospheric results favors $\nu_\mu \rightarrow
\nu_\tau$ oscillations.  Thus solar neutrinos may be the best tool 
for probing the new properties of the first-generation
$\nu_e$. \\

\noindent
{\it What is the significance of this subfield for nuclear physics
and science in general?}  \\
$\bullet$ The nuclei we study were created in stars and in stellar
explosions.  Solving the solar neutrino problem is the first step
in demonstrating we understand stellar evolution and nucleosynthesis quantitatively.
It opens the door to further studies in more explosive environments,
where nuclei exist in conditions not yet found in the laboratory. \\
$\bullet$ In the past two decades physics has made an extraordinary
investment in both accelerator and nonaccelerator experiments
to probe the standard model.  It now appears that the first sign
of new physics involves the neutrino.  Nuclear physicists started
the field of neutrino astrophysics with the chlorine experiment
and are now positioned to contribute to major discoveries in
particle physics. \\
$\bullet$ A knowledge of neutrino masses is crucial to the next
generation of precision cosmology experiments.  It is already
established that neutrinos are an important part of the universe's
mass, at least comparable to the visible stars.  Neutrino mixing
can alter the spectrum of cosmological neutrinos. \\
$\bullet$ Neutrino properties are crucial to understanding much
of nuclear astrophysics, including the supernova 
mechanism and the r-process. \\

\noindent
{\it What are the achievements of this subfield since the last
long range plan?} \\
$\bullet$ In 1995 SuperKamiokande was nearing completion.  It has
now produced results on the $^8$B neutrino spectrum of unprecedented
accuracy and found very strong evidence of atmospheric neutrino
oscillations. \\
$\bullet$ In 1995 SNO was under construction.  Today it is operating,
has surpassed its background goals, and has observed the $^8$B
neutrino spectrum. \\
$\bullet$ In 1995 no neutrino source of sufficient intensity was
available for measuring the responses of solar neutrino detectors.
GALLEX and SAGE have now been tested with
$^{51}$Cr neutrino sources, verifying the nuclear cross sections
and the efficiency of the chemistry. \\
$\bullet$ In 1995 Borexino's Counting Test Facility was under
construction.  The CTF experiment was successful, and construction
of the full detector is now well underway. \\
$\bullet$ KamLAND, an experiment that can directly probe part of
the neutrino oscillation parameter space relevant to solar neutrinos
and may also have $^7$Be neutrino detection capabilities,
is now under construction. \\
$\bullet$ In 1995 there were still reasonable suggestions for 
nonstandard solar models that could reduce the solar neutrino
discrepancy.  Today both the increasing precision of helioseismology
and the development of solar-model-independent neutrino analyses 
appear to rule out any such possibility. \\

\noindent
{\it What are the theoretical and experimental challenges facing 
the field?  Identify the new opportunities.} \\
$\bullet$ Definitive proof of oscillations must be obtained.  SNO's
ability to distinguish charged and neutral current events is the
outstanding opportunity to provide such proof. \\
$\bullet$ Various oscillation scenarios can account for the data,
some of which are quite difficult to distinguish unless low
energy solar neutrinos can be measured.  Can the US take the 
lead in developing and mounting one of the several promising 
experiments to measure the flux and flavor of these neutrinos? 
Among the proposals discussed by the working group were HERON,
HELLAZ, MOON, LENS, CLEAN, GaAs, and Cerenkov-triggered radiochemical
detectors employing Cl or I.  The ideas range from new
technology cryogenic detection schemes to hybrid detectors 
capable of simultaneously measuring solar neutrino reactions
and double beta decay. \\
$\bullet$ The atmospheric neutrino results are consistent with
maximal mixing, an unexpected result given the small mixing angles 
between quark generations.  Can theorists, aided by results from
SNO and other new experiments, find a compelling explanation
for the pattern of masses and mixing angles? \\
$\bullet$ Can detectors developed for solar neutrino research 
probe other neutrino sources: atmospheric neutrinos, geophysical
neutrinos, supernova neutrinos, and solar thermal neutrinos? \\

\noindent
{\it What are the resources required for this field?}\\
There must be continued strong support for the major experiments now
underway.  SNO appears to be functioning very well, but the 
continuation of that experiment through the neutral current phase
will require sustained effort.  Measuring the heavy-flavor
component of the solar neutrino flux is clearly the highest
priority.  Borexino and KamLAND are tackling the very challenging
problem of active detection of $^7$Be neutrinos.  While Italy
and Japan have the lead in these efforts, respectively, the
US participation is significant and must be continued.
     
But the principal challenge to US nuclear physics is to 
assume the lead in developing the next-generation active 
detector for the lowest energy solar neutrinos.  The US currently
lacks an effective mechanism for nurturing projects in the R\&D 
phase and for constructing new detectors once the concepts have
been proven.  This problem is long standing, and has led to
lost opportunities such as the gallium experiment.  The 
institutional support available elsewhere - Gran Sasso is the
outstanding example -- places the US at a disadvantage.  
Because we lack a facility like Gran Sasso to advocate for the subfield,
the community and the funding agencies must be more active 
in assessing a broad range of developing 
technologies; in distinguishing promising efforts from others,
strongly supporting those R\&D directions that make progress; 
and in mounting major experiments when the development stages 
have been completed. 

Judging from precedents like SNO and Borexino,
the typical scale of such major experiments is now in the \$25-50M
range.
  
\subsection{Neutrino Mass Working Group}

The neutrino mass working group focused on the status and future
of double beta decay experiments and tritium beta decay and other direct 
tests of neutrino mass. \\

\noindent
{\it What scientific questions is this subfield trying to answer?} \\
$\bullet$ Is lepton number conserved?  The most sensitive and most
direct test of this question is provided by neutrinoless double
beta decay. \\
$\bullet$ How does the neutrino transform under charge conjugation?
The neutrino, lacking any additive quantum numbers like electric charge, is 
unique among the known fermions in having an ambiguous behavior
under charge conjugation.  It may be its own antiparticle (Majorana)
or it may have a distinct antiparticle (Dirac).  The possibility of
both Majorana and Dirac masses is the key to the seesaw mechanism,
the most popular theory explaining why neutrinos are so much lighter
than their charged partners. \\
$\bullet$ What is the nature of neutrino mixing?  As a virtual 
process, neutrinoless double beta decay probes aspects of the 
neutrino mass matrix that, otherwise, are very difficult to test.
It is sensitive not only to very light Majorana masses ( below 1 eV),
but also to very heavy ones, above a TeV.  The mass derived from
double beta decay is sensitive to the relative CP eigenvalues of the 
mass eigenstates.  It is also sensitive to CP-violating phases in
the mass matrix. \\
$\bullet$ How does neutrinoless double beta decay probe new phenomena
beyond the standard model?  Again, as a virtual process, double
beta decay is particularly sensitive to new physics, even physics
residing at very high energies.  Examples include lepton-number-violating
right-handed couplings, Majorons, supersymmetry, ... \\
$\bullet$ What is the absolute scale of neutrino masses?  This is the
crucial question for cosmology and dark matter searches, yet cannot
be answered by either oscillation experiments, which depend on 
differences in the squares of the masses, or double beta decay,
where eigenstates with different CP eigenvalues intefere.
It can be measured in kinematic neutrino mass
experiments, with tritium beta decay being the outstanding example. \\

\noindent
{\it What is the significance of this subfield for nuclear physics 
and science in general?} \\
$\bullet$ Double beta decay -- neutrinoless and two neutrino -- is
a fundamental nuclear process.  It is the only open decay mode for
approximately 50 otherwise stable nuclei.  It is also the rarest
process yet measured in nature.  The basic decay process involves
a two-nucleon correlation and a nuclear polarizability, and thus 
is fascinating from the perspective of nuclear structure theory. \\
$\bullet$ Opportunities to study second-order weak decays in nature
are extremely rare.  Double beta decay is one of only two such
possibilities in particle physics. \\
$\bullet$ The question of lepton number violation in the early
universe is crucial to cosmology.  It is connected, in the standard
model, to possible mechanisms for baryogenesis.  Early universe lepton number
asymmetries can trigger oscillations that distort the neutrino
distributions, producing warm -- not hot -- neutrino dark matter. \\
$\bullet$ The question of the absolute scale of neutrino masses 
is crucial to dark matter studies, including interpretations of
the cosmic microwave background and large scale structure.  
Tritium beta decay is a direct test of this mass scale. \\
$\bullet$ Double beta decay (and solar neutrino) experiments are technologically relevant.  Low
level counting and ultrapure materials have industrial significance. \\
  
\noindent
{\it What are the achievements of this subfield since the last
long range plan?} \\
$\bullet$ Thirty years of effort was required before the allowed
process, two-neutrino double beta decay, was observed in 1987.
Today accurate lifetimes are known for approximately 12 nuclei. \\
$\bullet$ Extraordinary efforts to reduce backgrounds has resulted
in a ``Moore's law'' for neutrinoless double beta decay: 
over the past two decades, lifetime limits have improved by a
factor of two every two years.  The current limit on the 
Majorana mass is in the range (0.4-1.0) eV, with the spread
reflecting nuclear matrix element uncertainties. \\
$\bullet$ Detector technology has greatly improved in areas
such as backgrounds, detector mass, the use of isotopically 
enriched sources, and cryogenics. \\
$\bullet$ Double beta decay theory has improved significantly.  Shell model methods
have been develop to treat the intermediate nuclear Green's
function in two-neutrino decay.  Full or nearly full fp-shell
diagonalizations have been done.  Shell model Monte Carlo
methods have been developed and checked against exact shell
model results.  Virtually all of this work has occurred since
the last long range plan. \\
$\bullet$ Tritium $\beta$ decay mass limits have reached 
2.2 eV (95\% c.l.), a bound important to cosmology. \\
$\bullet$ At the time of the last Long Range Plan, the observation 
of excess events near the endpoint affected the field's confidence in
tritium $\beta$ decay mass limits.  Recently the Mainz group has traced much
of the effect to energy losses due to rough source surfaces.
Their latest results (98-99) appear to be free of any problems. \\

\noindent
{\it What are the theoretical and experimental challenges facing
the field?  Identify the new opportunities.} \\
$\bullet$ Radiogenic and cosmogenic backgrounds have been 
tremendously surpressed by new techniques for refining ultrapure
materials and by mounting experiments underground.  The most
challenging background in many cases is now the high-energy tail
of the $2\nu$ process, a serious limitation for detectors
lacking excellent electron energy resolution.  Thus high
resolution detectors must be developed. \\
$\bullet$ With lifetime limits now above $10^{25}$ years, the
next generation of detectors must employ larger masses to make
progress.  The counting rate is a fundamental limit at current
$\sim$ 10 kg detector masses.  Thus much larger 
masses ( $\sim$ 1000 kg) are being planned in future experiments. \\
$\bullet$ The physically relevant scales for $0\nu$ double beta decay
experiments are still unclear.  Current theoretical models
that explain the solar and atmospheric neutrino results with
Majorana neutrinos predict a broad range of double beta decay
masses (typically from 1 eV to $10^{-5}$ eV). \\
$\bullet$ The goal of future direct $\nu_e$ mass
searches is a sensitivity below 1 eV. The new ideas include the proposed
7m Karlsruhe spectrometer and a cryogenic calorimeter using Re. \\

\noindent
{\it What are the resources required for this field?} \\
The current generation of neutrinoless double beta decay experiments includes the Heidelberg-Moscow
and IGEX experiments on $^{76}$Ge, the Caltech-Neuchatel effort on
$^{136}$Xe, and the ELEGANTS and NEMO-3 $^{100}$Mo 
measurements.  They have comparable goals (lifetime limits of 
$\sim 10^{25}$ years) and typically involve parent isotope masses
of $\sim$ 10kg.  The Heidelberg-Moscow experiment, which has
acquired more than 35 kg-years of data, has set a limit of 
$2 \times 10^{25}$ years on the $^{76}$Ge lifetime.
All of these experiments are being conducted outside the US, though
several involve US collaborators.

The new large-mass proposals have as their goal Majorana mass
limits in the range of 0.03-0.10 eV.  This is an important goal
since the $\delta m^2$ deduced by SuperKamiokande, our first
indication of the scale of neutrino masses, is centered around $\sim (0.05$ eV)$^2$.
There are also strong claims from cosmologists that, once results
have been obtained from MAP, PLANCK, and the Sloan Digital Sky
Survey, knowledge of neutrino masses in the 0.1-1.0 eV range 
will be important to their analyses.  The proposed experiments
include MAJORANA and GENIUS (enriched $^{76}$Ge), CUORE (cryogenic detector
using $^{130}$Te), MOON ($^{100}$Mo foils with plastic scintillator),
EXO (a laser tagged TPC using $^{136}$Xe), and CAMEO ($^{116}$Cd and
$^{100}$Mo in Borexino's CTF).  Some of these detectors are very
well developed, while others require considerable R\&D.
Some of the proposals, such as MAJORANA, GENIUS, MOON, and EXO, depend on Russian 
isotopic enrichment facilities which are currently available, but
may not be so indefinitely.

Among the resources needed are consistent support for R\&D in
those cases where significant development is necessary; 
agency help in defining procedures where developed projects
can be evaluated and supported; and support for international
collaborations (most of the next-generation experiments listed
above are international).  The anticipated cost of a
typical 1000 kg experiment is $\sim$ \$10M, exclusive of
isotope enrichment costs (which may be provided by other agencies).
There is need for some redundancy in double beta decay studies because
of nuclear matrix element uncertainties and because the ultimate
sensitivity of new approaches is often difficult to predict.

A US site for mounting double beta decay experiments is another issue.
While the needs of experiments differ, current experiments 
typically require about 2000 m.w.e. coverage.  Thus near-term
requirements can be satisfied by sites like WIPP and the Soudan Mine, but deeper sites may
be required for some next-generation detectors.  Another issue
is cosmic ray induced activities: activities such as $^{68}$Ge
must be allowed to decay away.  Thus underground storage of 
materials, in anticipation of future experiments, is under 
discussion.

Dual-purpose detectors for solar neutrino and double beta decay
are another interesting possibility in the next decade.  MOON proponents
anticipate solar neutrino rates for $pp$ and $^7$Be neutrinos
comparable to their neutrinoless $\beta \beta$ decay rate goals.
The problems confronting both types of measurements, small
rates and troublesome backgrounds, can be solved with 
highly instrumented detectors that exploit coincidence techniques
to isolate the signals of interest.

While shell model treatments of double beta decay have become
more sophisticated in the last ten years, fundamental issues still
need attention.  Probably the most important is the effect
of shells in the excluded space: how do these renormalize the
shell-model $\beta \beta$ decay operators?  The theory is not 
an experimental show-stopper -- any observation of 0$\nu$ $\beta \beta$ decay demonstrates lepton
number violation -- but is important in translating lifetime limits
into upper bounds on neutrino masses.

In tritium $\beta$ decay near-term activity will focus on the
Karlsruhe-Mainz-Troitsk project, an effort to push mass limits to $\sim$ 0.5 eV
with a massive 7m spectrometer.  The estimated cost is 
\$10-15M.  Other groups 
have been invited to join.  Thus 
support is needed for US collaborators wanting to help in this effort.
There are also interesting cryogenic Re calorimeters
under development in Genova and Milano.  On the longer term,
very severe obstacles will have to be overcome to 
further increase sensitivies to $\sim$ 0.1 eV.  Molecular excited state 
contributions are one of the very troublesome issues at this
level.

\subsection{Supernova Neutrinos and the Supernova Mechanism Working Group}
This working group focused on the theoretical challenge of building
realistic models of core-collapse supernovae, the experimental
challenge of building and operating neutrino observatories to
measure the flux and flavors of neutrinos from the next galactic
supernova, and related astrophysics issues. \\

\noindent
{\it What scientific questions is this subfield trying to answer?} \\
$\bullet$ What is the mechanism by which a core-collapse supernova
ejects its mantle?  Can we build a quantitative standard model
of the explosion, including neutrino production and associated
nucleosynthesis, such as the r-process and the neutrino process?\\
$\bullet$ What experiments can be done to test such a standard 
model?  Can we use the nucleosynthesis, particularly the pattern
of r-process metals, to diagnose the explosion, in analogy with
the use of d, $^{3,4}$He, and $^{6,7}$Li to test the big bang?
Can we use the neutrino flux from the next galactic supernova
to learn about the explosion mechanism and, possibly, to probe
properties of the protoneutron star?  Can we measure the gravitational
wave signal in LIGO, and supernova gamma rays in INTEGRAL? \\
$\bullet$ Can we exploit supernovae to search for new phenomena,
including neutrino oscillations and neutrino masses? \\ 
  
The supernova mechanism is one of the outstanding challenges
in nuclear theory and theoretical astrophysics, involving an extraordinary range of physics.
To specify the initial conditions for the explosion the 
massive progenitor star must be evolved through its various 
burning stages, to formation of the inert iron core.
This problem couples laboratory nuclear astrophysics -- 
including open problems like the $^{12}$C + $\alpha$ S-factor --
with stellar evolution, and is very much an extension of the
program that began with the solar neutrino problem.  The
description of the core bounce requires us to predict the
behavior of bulk nuclear matter at densities and temperatures
not otherwise accessible.  New phenomena -- mixed or quark-matter
phases, color superconductivity, kaon condensation -- could 
affect the equation of state.  Both the early deleptonization
of the star and the subsequent cooling require a detailed
treatment of neutrino transport through the nuclear medium,
and an understanding of the various processes that determine the opacity.
Shock wave propagation through nuclear matter must be understood.
The nucleosynthesis depends on relationships between the 
explosion dynamics, the neutrino physics, and laboratory
astrophysics.  The explosion determines the timescale
for the nucleosynthesis.  Neutrino reactions control the isospin
of the nuclear matter.  Laboratory astrophysics must determine
the masses and the $\beta$ decay lifetimes important to 
the r-process and other explosive nucleosynthesis.

The neutrino fluxes produced by a supernova provide unique 
opportunities to learn about neutrino properties.  As the 
neutrinosphere resides at a density $\sim 10^{12}$ g/cm$^3$,
supernovae allow us to extend our tests of matter effects
on oscillations by 10 orders of magnitude.  Thus MSW effects,
even for very small mixing angles of $10^{-5}$, can distort
the neutrino spectra.  The entire range of cosmologically
interesting masses can be probed in this way.  In particular,
supernovae may provide our best laboratory for investigating
$\nu_e - \nu_\tau$ oscillations.  Kinematic tests of neutrino
mass can be made by studying arrival times on earth as a 
function of flavor or energy.  In this way it may be 
possible to greatly reduce mass limits for the $\nu_\tau$
and $\nu_\mu$. \\

\noindent
{\it What is the significance of this subfield for nuclear physics
and science in general?} \\
$\bullet$ Supernovae are thought to have produced about half
of the heavy nuclei found in nature.  Nucleosynthesis is a central
question for nuclear physics. \\
$\bullet$ To the extent that we can understand such synthesis,
we can predict, given a galactic model, how metallicities
evolve.  This opens up a wonderful intersection with 
astronomy, including both abundance determinations and gamma
ray astronomy. \\
$\bullet$ Neutron stars are the only example in nature of the
nuclear theorist's test case, bulk nuclear matter.  It is very
likely that new phenomena exist at neutron star densities.
In the next decade precise mass/radii determinations are likely
to be made.  This will provide a crucial check on our theories
of the equation of state of dense nuclear matter. \\
$\bullet$ Core collapse supernovae (and neutron star merges)
may produce detectable gravitational radiation.  Accurate 
modeling of the collapse could help LIGO experimentalists by
defining the wave forms that they must find. \\
$\bullet$ Supernova neutrino detection is a key part of the 
``supernova watch'' program that also involves gravitational
wave detectors and optical observatories. \\
$\bullet$ Supernova modeling is a terascale (and beyond) ``grand challenge''
problem that requires collaboration between nuclear theorists, astrophysicists,
and computer scientists.  Many of the underlying issues,
such as radiation transport, hydrodynamics, shock wave propagation,
and the mathematical challenge of scalable algorithms for
large, sparse, linear systems, are common to problems ranging
from medical imaging to climate prediction to internal combustion.
Thus the developments from supernova models will
benefit many other sciences. \\ 

\noindent
{\it What are the achievements of this subfield since the last long range plan?} \\
$\bullet$ At the time of supernova 1987A, two neutrino detectors 
were operational and $\sim$ 18 events were recorded.  Today there 
are four operating detectors and three others that should be
operational in the next 1-2 years.  Approximately $10^4$ neutrinos
should be counted at the time of the next galactic supernova. \\
$\bullet$ The first semi-realistic two-dimensional simulations of
supernova explosions have been performed.  This could be an important step
in understanding the mixing apparent in the ejecta of
observed supernovae.\\
$\bullet$ Full Boltzmann neutrino transport has been implemented in
one dimensional models. \\
$\bullet$ Significant progress has been made in descriptions of the
progenitor, e.g., multi-D models that account for convection and
rotation.  Improved electron capture and beta decay rates and
improved neutrino opacities have made the input microphysics much
more realistic. \\
$\bullet$ Progress has been made in modeling the r-process, including
improved weak interaction rates, a better understanding of the 
effects of mass formula uncertainties and phenomena such as the vanishing of
shell closures, and inclusion of neutrino postprocessing effects. \\
  
\noindent
{\it What are the theoretical and experimental challenges facing
the field?  Identify the new opportunities.} \\
$\bullet$ The key theoretical challenge is to develop a 
supernova standard model that incorporates realistic neutrino transport
and microphysics.  Current 1D models generally fail to explode. 
This could reflect some flaw in our understanding of the
physics, or the importance of doing multi-D simulations. \\
$\bullet$ Test relevant microphysics input into supernova simulations,
such as mass formulas used in r-process synthesis and neutrino-nucleus
cross sections important to opacities and nucleosynthesis, by
direct laboratory measurements at RIA, ORLAND, and other facilities. \\
$\bullet$ Test supernova models by comparing predicted supernova
neutrino flavor, energy, and time distributions to measurements
made in underground neutrino detectors. \\
$\bullet$ Exploit the next galactic supernova to make kinematic 
tests of neutrino masses.  More significant results can be 
obtained if
sharp temporal features in the neutrino flux -- the rise time
or a sudden termination of $\nu$ emission due to black hole
formation -- can be identified. \\
$\bullet$ Exploit the next galactic supernova to search for 
the MSW flavor transformation $\nu_\tau \rightarrow \nu_e$.
Because supernova $\nu_e$s are more strongly coupled to the
matter, they are predicted to be substantially less energetic 
than the heavy flavor neutrinos.  Thus such a MSW oscillation 
will produce a distinctive spectral inversion, distorting 
the angular distribution of events in detectors like SuperKamiokande. \\

There are many open questions in supernova modeling that could be
addressed by a ``grand challenge'' effort.  A variety of physics --
neutrino heating, convection, rotation, magnetic fields, general
relativity -- are inadequately modeled in current multi-D
simulations.  It is not known which of these effects may be
essential to successful explosions.  Nor is it clear how dependent
(or independent) explosions may be on the class of progenitor
star. \\
  
\noindent
{\it What are the resources required for this field?} \\
One of the fundamental difficulties for the field is the low rate 
of galactic supernovae, estimated to be $\sim$ 1/30 years.  This
corresponds to a timescale that exceeds some (though not all)
neutrino detector lifetimes.  The challenge, then, is to begin
to view neutrino detectors as observatories, rather than 
experiments.  Some compromises may be required because highly
instrumented, high maintenance detectors become more costly,
both in dollars and in the human investment, when operated over
decades.  Thus there is some merit to arguments that
low-tech, flavor-specific experiments, such as the radiochemical detectors Cl, I, 
and SAGE/GNO, have a role to play.

When possible, it is clearly preferable to exploit detectors 
built for other purposes -- SuperKamiokande, SNO, or a next-generation
proton decay detector like UNO -- as supernova observatories.
This allows the physicists involved with the detector to do other
physics while waiting for a rare supernova.  Yet there are
proposals for dedicated experiments, such as OMNIS, that  
are designed to minimize manpower requirements.

If supernova observatories are exclusively multipurpose detectors,
then in some sense they monitor the galaxy for free.  But to
ignore the supernova physics in designing and operating such
detectors is clearly a mistake.  It is essential that detectors
with the requisite capabilities monitor the galaxy at all times,
to avoid missing a once-in-a-lifetime opportunity.  This is a
theme of the supernova watch.  In the case of SN1987A, we measured
supernova $\bar{\nu}_e$s.  The goal, at the time of the next
supernova, should be to measure separately the properties of
the $\nu_e$, $\bar{\nu}_e$, and heavy flavor fluxes.
Water Cerenkov detectors have excellent capabilities for 
$\bar{\nu}_e$s; the charge current reaction in SNO provides a
clean signal for $\nu_e$s.  But SNO may
operate for only a decade, and neither SNO nor SuperKamiokande has a 100\% duty
cycle.  While there are strategies for extracting a neutral current
(and thus dominately heavy flavor) signal from SuperKamiokande,
the deuteron breakup reaction in SNO, which will produce about
800 events for a supernova at the galactic center, appears to
be the better monitor of this flux.  Scintillation detectors, such  
as KamLAND and Borexino, are also interesting neutral current
detectors because neutrinos will excite the 15.11 MeV M1
transition in $^{12}$C.

Clearly the nuclear physics community needs to be highly involved
in supernova watch plans.  Decisions to turn off detectors must 
take into consideration whether supernova capabilities are being
lost.  This also applies to scheduled maintenance.

The arguments for a theory initiative in supernova physics are
very strong.  This modeling is central not only to neutrino physics,
but also to other major nuclear physics
initiatives, such as RIA.
The development of multi-D models with realistic neutrino transport and
microphysics is possible at this time.  Presuming that 
terascale machines are made available, the primary resource 
needed is person power: the groups currently involved in supernova
theory are greatly understaffed.
A reasonable starting budget for such an
initiative is \$2.0M/year, most of which should be invested in
young scientists who would attack 
the neutrino transport, hydrodynamics, and computer science issues
associated with supernova modeling, as well as critical issues
involving the underlying microphysics, such as the nuclear structure
important to neutrino-nucleus scattering and other weak interactions,
the nuclear equation of state at high density, and neutrino
opacities.

\subsection{Underground Laboratories Working Group}
The underground laboratories working group considered not only
underground sites, but also interdisciplinary experiments, such
as those on atmospheric neutrinos or neutrinos produced by 
high-energy astrophysical sources, conducted in such sites. \\

\noindent
{\it What scientific questions is this subfield trying to answer?} \\
$\bullet$ What type of environment, isolated from both
cosmic ray and natural radioactivity backgrounds, can be provided 
to optimize the success of future background-sensitive experiments?
How should such a facility (or facilities) be operated to meet
the needs not only of nuclear physics, but of physics and science
in general? \\
$\bullet$ What can be learned by extending the program of
astrophysical neutrino detection to higher energies?  In analogy
with solar neutrinos and supernova neutrinos, could such a 
program allow us to probe the structure of active galactic nuclei
and other high-energy objects? \\
$\bullet$ What contributions can nuclear physics make to the
atmospheric neutrino problem and to proton decay and other searches
for physics beyond the standard model? \\

\noindent
{\it What is the significance of this subfield for nuclear physics
and science in general?} \\
$\bullet$ Despite early leadership in the field of underground science,
the US has fallen behind Europe and Japan in providing facilities
for such experiments.  The US community is largely engaged in
overseas projects.  The few efforts within the US, such as the
Homestake solar neutrino program, manage in active mines.   
The shortage of suitable underground sites is a concern for
proton decay, dark matter, and similar searches for new physics,
projects important to our colleagues in particle and astrophysics. \\
$\bullet$ Atmospheric neutrinos are another tool to probe the 
physics of neutrino mass and mixing of interest to the nuclear
physics community.  SNO will have significant capabilities
for atmospheric neutrinos.  As the source distance varies from
the height of the atmosphere to the diameter of the earth,
very clean tests of oscillations can be made with atmospheric
neutrinos if the oscillation length lies in this range. \\
$\bullet$ Many phenomena in astrophysics -- such as $10^{21}$ eV
cosmic rays and gamma ray bursts corresponding to isotropic
sources of energy $10^{53}$ ergs -- involve extraordinary 
scales of energy and particle acceleration.  Even in our own
galaxy there are hints, from AGASA, of $10^{18}$ eV events.
The detection of neutrinos produced by such natural accelerators
might help us understand the acceleration mechanism and pinpoint
the source.  Nuclear physics can contribute to such high-energy
astrophysics questions because of our interest in water
Cerenkov and other neutrino detection schemes. \\

\noindent
{\it What are the achievements of this subfield since the last 
long range plan?} \\
$\bullet$ SNO has been constructed and has surpassed its background
specifications, demonstrating that a clean-room environment can
be maintained at great depth, even in an active mine. \\
$\bullet$ WIPP, the waste isolation facility in New Mexico, has
offered to host scientific experiments.  This provides a 
US laboratory site at moderate depth ($\sim$ 2000 m.w.e.).
The depth is comparable to that of the Soudan Mine, but 
access is easier.  Because it is located in a salt formation, 
U and Th background levels are low. \\
$\bullet$ SuperKamiokande has measured with very good statistics a
distinctive zenith angle dependence in the ratio of electron to
muon events.  There are very strong arguments attributing 
this to $\nu_\mu \rightarrow \nu_\tau$ oscillations with
a nearly maximal mixing angle.  Most experts accept this result
as the first demonstration of physics (nonzero neutrino masses,
flavor mixing) beyond the standard model. \\
$\bullet$ AMANDA, the high-energy neutrino detector located  
1500-2000m below the surface of the Antarctic ice sheet, 
was commissioned in February, 1997.  The experimentalists have
observed atmospheric neutrinos and are searching for astronomical
sources. \\

\noindent
{\it What are the theoretical and experimental challenges facing
the field?  Identify the new opportunities.} \\
$\bullet$ There are outstanding opportunties to create a deep
underground national laboratory that will serve the next generation
of solar neutrino, double beta decay, dark matter, atmospheric neutrino, and proton
decay experiments.  Deep sites are also important to accelerator
measurements of astrophysical S-factors, and potentially interesting
for other sciences and industry.

The model for such a national laboratory is Gran Sasso, located
at an average depth of 4300 m.w.e. and with horizontal access off a
highway excavated through the Gran Sasso d'Italia.
The laboratory has been in existence since the early 1980's.
Gran Sasso comprises 3 halls ($\sim$ 100 m $\times$ 18m $\times$
18m), external offices, a computing center, technical and
engineering services, electronics and chemical laboratories,
a machine shop, library, conference room, and stockroom.
The competion for space is keen.  The laboratory currently hosts
a broad program of experiments: the GNO successor to the GALLEX
solar neutrino experiment; Borexino; the dark matter search
DAMA; the Heidelberg-Moscow $^{76}$Ge double beta decay experiment;
the EASTOP air shower array for cosmic ray physics (and for
coincidence with underground detectors), located on top of the
mountain; the ICARUS liquid Ar detector; LVD, a 1.6 kton 
liquid scintillator detector; MACRO, a large monopole
detector; and LUNA, a low-energy accelerator for nuclear astrophysics.  

The Kamioka laboratory, located in a mine in the Japanese alps,
is also becoming a multipurpose facility.  Activities include SuperKamiokande, KamLAND,
a gravity wave detector under construction, and double beta
decay.

There has been serious discussion in the US of a deep underground
national laboratory since the early 1980s.  The question has
become very urgent with the announcement that the Homestake
Mine, in South Dakota, will close in 16 months.  Homestake is
a deep, hardrock mine with a large shaft (15 $\times$ 20 ft)
running to 4850 ft; additional levels exist every 150 ft, to
8000 ft.  The State of South Dakota, in combination with the
South Dakota School of Mines, has expressed interest in taking
on the operations, management, and liability burdens that would
be associated with an underground laboratory.  The mine has
considerable infrastructure (pumps, power, air exhaust systems, 
multiple shafts).  But significant investments are needed to
produce an above-ground campus comparable to that at Gran Sasso;
to install modern lifts that utilize the full dimensions of the
shafts; to produce large halls of the type existing at Gran
Sasso; and to engineer areas for cryogenics and other facilities
where safety is a concern.

There is also a proposal for constructing a horizontal access
laboratory by tunneling beneath Mt. San Jacinto, near Palm
Springs.  A laboratory located at the end of a 2.5 mile tunnel
would provide 6000 ft of rock overbunden.  Although this
requires construction of a laboratory and its infrastructure
from scratch, the plan offers the advantages of horizontal
access and proximity to a number of physics 
laboratories in California. 
  
The Sudbury Neutrino Observatory is a third possible  
deep site in North America.  Though an active mine, the
success of SNO contruction demonstrates that science requiring
a clean-room environment can be done there.  The SNO site
is quite deep, 2039 m.

These possibilities for deep sites,
together with the existing shallower sites at the Soudan Mine
and WIPP, should be the starting point for a community discussion
of how to prepare for the next generation of underground
experiments.  These sites have complementary aspects: different
radioactivities, access, depth potentials, etc.  The 
community has an opportunity to consider which facility or
combination of facilities will help the next generation of
experiments reach their potential.  As Gran Sasso has proved, both the
underground site and the supporting infrastructure are 
important in facilitating new experiments. \\
$\bullet$ The successful commissioning of AMANDA opens up the
possibility of large Antarctic arrays to do high energy neutrino
astronomy.  AMANDA has given the US leadership in this area.
The next generation detector, ICECUBE, is designed to map
the neutrino sky from GeV to PeV energies, determining both
the diffuse flux from galactic and extra-galactic sources
and point sources, such as active galactic nuclei or gamma
ray bursters.  High energy neutrinos are unique tracers of
high energy protons and nuclei that we know are accelerated
to extraordinary energies somewhere in the cosmos.  The behavior 
of nuclei at very high energies and their interactions with
the interstellar medium are topics of interest to nuclear physicists. \\

\noindent
{\it What are the resources required for this field?} \\
The creation of a national deep underground laboratory is a
major investment, the largest discussed in this report.  
In the case of Homestake, there is a large investment already
made by the miners: the value of the existing mile-long 15 $\times$ 20 ft
shaft is considerably in excess of \$100M.  The additional investment
that will be needed from scientific agencies to convert
Homestake into a suitable national facility may be smaller,
but is still significant.  The costs include improved lifts,
the experimental halls, and the above-ground facilities of
the type provided by Gran Sasso.  The cost of the experimental program of
such a facility, extrapolating from Gran Sasso, is likely
in the \$20-25M/year range.  There are important 
efficiencies in such a laboratory because experiments can make
use of a common infrastructure.

The construction costs of an ab initio laboratory like San Jacinto are more
difficult to estimate.  A reasonable extrapolation of the 
estimates made in the early 1980s, when the proposal was 
first discussed, yields $\sim$ \$100M.

The construction and operations costs of such a facility 
presumably would be shared between nuclear and particle physics:
dark matter searches and proton decay experiments are among
the candidate experiments requiring significant cover.
It is unlikely that the use of such a laboratory would be
confined to nuclear and particle physics: isolated environments
are also of interest to geophysicists, the electronics industry,
biologists, and gravity wave experimentalists. 

\subsection{Reactor and Accelerator Neutrinos} 
{\it What scientific questions is this subfield trying to answer?}\\
$\bullet$ Can neutrino oscillations be observed under controlled 
laboratory conditions? \\
$\bullet$ Can neutrinos provide new information on the structure
of nucleons and nuclei, such as strangeness content? \\
$\bullet$ Can laboratory neutrino sources be exploited to test our
understanding of neutrino-nucleus cross sections important to
supernovae and to the solar neutrino problem? \\

\noindent
{\it What is the significance of this subfield for nuclear physics
and for science in general?} \\
$\bullet$ Despite the strong evidence for neutrino oscillations 
from atmospheric and solar neutrino studies, the underlying 
physics issues are so important that confirmation of oscillations
in the laboratory is crucial.  The use of known neutrino
sources and the ability to adjust the source-target distance
are among the advantages of accelerator and reactor neutrinos.
Disappearance and appearance measurements can be made. \\
$\bullet$ Neutrinos are potentially interesting as probes of
strangeness in the nucleon and nucleus, with simpler radiative
corrections.  If some of the suggested experiments can be done,
the results complement similar studies done at JLab and elsewhere.\\
$\bullet$ There are very few quantitative tests of the 
accuracy of calculated neutrino-nucleus cross sections.  The
renormalization of the effective shell model axial vector
coupling $g_A$ is known from $\beta$ decay, while muon capture
probes first-forbidden weak responses for time-like four-momenta.
But apart from these constraints,
most of the nuclear physics used in describing supernova
neutrino-nucleus cross sections (space-like four-momentum 
transfers, important allowed and first forbidden transitions)
has not been subjected to detailed experimental tests.
Yet many aspects of supernova physics, including nucleosynthesis,
require accurate cross sections. \\

\noindent
{\it What are the achievements of this subfield since the last 
long range plan?} \\
$\bullet$ LSND was completed in 1998, and KARMEN II has reported three years of 
data (2/97-3/00).  Evidence for $\bar{\nu}_\mu \rightarrow \bar{\nu}_e$ 
has been found in the LSND experiment.  KARMEN has found no
evidence for oscillations, though a portion of the
LSND allowed range, corresponding to small mixing angles and
masses, is not ruled out. \\
$\bullet$ The Chooz and Palo Verde reactor $\bar{\nu}_e$ oscillation
experiments constrained $\delta m^2$ to below $10^{-3}$ eV$^2$ in
the disappearance channel at maximum mixing angle.  This became
an important constraint on the interpretation of the SuperKamiokande
atmospheric neutrino results. \\
$\bullet$ Early results from the K2K long-baseline oscillation 
experiment disfavor the no oscillation hypothesis at about the
two standard deviation level.  This is the first hint that the
SuperKamiokande atmospheric neutrino results may have laboratory
confirmation. \\
$\bullet$ LSND and KARMEN obtained $^{12}$C exclusive charge 
current cross sections for exciting the ground state of $^{12}$N.
The results are in good agreement with theory.  Inclusive
charge current cross sections were also obtained.  KARMEN 
measured neutral current neutrino excitation of the 15.11 MeV
state in $^{12}$C.  These results are the first obtained for 
complex nuclei. \\
$\bullet$ FermiLab's MiniBooNE, the followup experiment to 
LSND, and KamLAND, the first reactor/accelerator neutrino experiment to
address part of the oscillation parameter space relevant to solar
neutrinos, are under construction. \\

\noindent
{\it What are the theoretical and experimental challenges facing
the field?  Identify the new opportunities.} \\
$\bullet$ The completion of MiniBooNE is very important.  If 
both the atmospheric and solar neutrino problems are attributed
to neutrino oscillations, and if the LSND results are correct,
then a good fit to the data can only be obtained by hypothesizing
a fourth light neutrino with sterile interactions.  Thus confirming
or ruling out the LSND results has important consequences for
interpretations of the neutrino data. 
If the LSND results are confirmed by MiniBooNE, it will be important
to build a second detector at a different distance in order to
define the oscillation parameters precisely. \\
$\bullet$ The KamLAND reactor antineutrino experiment now under
construction will break exciting new ground: it will
be the first laboratory
experiment to probe $\delta m^2$ values directly relevant to
the solar neutrino problem.  The projected sensitivity covers
all of the large mixing angle solar neutrino solution.  Thus
completion of this experiment must be a very high priority for nuclear
physics. \\
$\bullet$ Similarly, accelerator oscillation experiments testing
the atmospheric neutrino parameter space are crucial.
K2K, which already has very interesting data, and MINOS must
be completed. \\
$\bullet$ The Spallation Neutron Source now under construction 
at Oak Ridge will, as a byproduct of operations, produce an 
intense source of neutrinos with a pulsed time structure
(similar to that of ISIS/KARMEN) and with a very favorable
$\bar{\nu}_e / \bar{\nu}_\mu \sim 3 \times 10^{-4}$ ratio.
This will be the most intense, pulsed, intermediate energy
neutrino source available.  An interesting program of possible
experiments has been discussed, some of which exploit the
similarities between the SNS neutrino spectrum and that from
a supernova.  The opportunity to build a neutrino bunker -- a
shielded room in which experiments can be mounted -- should be
seized.  It is important to situate the room as close as possible
to the SNS mercury beam stop.  This facility (ORLAND) will stimulate
the community to propose detectors and experiments: the 
possibilities include oscillation experiments, searches for 
isoscalar axial charge transitions, and the continuation of the
neutrino-nucleus cross section program begun by KARMEN and
LSND. \\
$\bullet$ Much of the underlying physics of ORLAND, the supernova
mechanism, and other problems discussed here involves nuclear
structure theory.  The US nuclear theory program is currently quite
weak in this area.  Neutrino physics has strong student appeal
and, because it involves many nuclear structure issues, provides
an opportunity for training students in an area of some
national importance. \\

\noindent
{\it What are the resources required for this field?} \\
It is important to continue nuclear physics support for the 
MiniBooNE and KamLAND efforts.  These experiments focus directly
on issues of importance to nuclear physics: checking
the LSND claims and probing neutrino oscillation parameters
relevant to the solar neutrino problem.

The SNS bunker requires a significant investment, perhaps 
\$15M.  The additional cost of detectors for the
experimental program has been estimated to be $\sim$ \$45M.

Two of the long range plan initiatives now under consideration --
the neutrino program outlined here and the Rare Isotope 
Accelerator -- have important links to nuclear structure theory.
There are very few US nuclear structure theorists of 
age $\lsim$ 40 years occupying tenure track university or national
laboratory positions.  (Interestingly, those few all seem to have close 
connections to weak interactions and neutrino physics, 
particularly neutrino astrophysics.)  It is important,
for the success of the experimental part of the LRP program,
to enhance the nuclear theory program in the relevant areas
of nuclear structure and nuclear astrophysics.
The creation of 20 entry-level university nuclear theory/nuclear astrophysics
positions over a period of 5 to 10 years would require an
increase in the theory budget of about \$3M/year.  
This \$3M investment could be used initially to fund tenure-track bridge
positions, then gradually rolled over to provide
continuing research support (summer salary, graduate students,
and postdocs) once the bridges are completed.
    
\section{Acknowledgements}
The Seattle Neutrino Workshop was hosted by the University of 
Washington and cosponsored by the UW's Institute for Nuclear
Theory and Center for Experimental Nuclear Physics and
Astrophysics, and by Oak Ridge National Laboratory.

The success of the workshop is due to the enthusiastic participation
of the neutrino physics community.  The talks presented at the
workshop can be found at
\begin{center}
int.phys.washington.edu
\end{center}
(click on talks online, then on Neutrino Workshop).
The organizing committee, working group convenors, and workshop
participants are listed in the following pages.

\pagebreak
\begin{center}
ORGANIZING COMMITTEE \\
\vspace{.7cm}

John Bahcall, IAS, Princeton \\
A. Baha Balantekin, Wisconsin \\
Stuart Freedman, Berkeley \& LBNL \\
Wick Haxton, Washington (cochair) \\
Kevin Lesko, LBNL \\
Hamish Robertson, Washington (cochair) \\
\end{center}
\vspace{0.3cm}

\begin{center}
WORKING GROUPS AND CONVENORS \\
\vspace{.5cm}

Solar Neutrinos \\
A. Baha Balantekin (Wisconsin) and Bob Lanou (Brown) \\

\vspace{.5cm}
Supernova Neutrinos, the Supernova Mechanism, and Related Astrophysics \\
George Fuller (UCSD), Ken Lande (Penn), and Tony Mezzacappa (ORNL) \\

\vspace{.5cm}
Reactor and Accelerator Neutrinos \\
Frank Avignone (South Carolina), Bill Louis (LANL), and Petr Vogel (Caltech) \\

\vspace{.5cm}
Underground Laboratories (including atmospheric and high energy neutrinos) \\
Todd Haines (LANL) and Kevin Lesko (LBNL) \\

\vspace{.5cm}
Neutrino Mass \\
Wick Haxton (Washington) and John Wilkerson (Washington) \\
\end{center}

\pagebreak

\begin{center}
LIST OF PARTICIPANTS \\
\vspace{.7cm}

Aalseth, Craig, University of South Carolina \\
Adelberger, Eric, University of Washington \\
Arnett, David, Steward Observatory \\
Avignone, Frank T., University of South Carolina \\
Balantekin, Baha, University of Wisconsin \\
Barger, Vernon, University of Wisconsin \\
Beacom, John, FermiLab \\
Beene, James, Oak Ridge National Laboratory \\
Bernstein, Robert, FermiLab \\
Bertrand, Fred, Oak Ridge National Laboratory \\
Blondin, John, North Carolina State University \\
Bonvicini, Giovanni, Wayne State University \\
Bowles, Thomas J., Los Alamos National Laboratory \\
Boyd, Richard N., Ohio State University \\
Brice, Steve, Los Alamos National Laboratory \\
Bruenn, Stephen, Florida Atlantic University \\
Bullard, Theresa, University of Washington \\
Butler, Malcolm, Saint Mary's University \\
Calaprice, Frank, Princeton University \\
Cardall, Christian, UT Knoxville/ORNL \\
Carter, Ken, Oak Ridge National Laboratory \\
Chen, Mark, Queen's University \\
Cline, David B., UCLA \\
Corey, Robert, SD School of Mines \\
De Braeckeleer, Ludwig, Duke University \\
Dean, David, Oak Ridge National Laboratory \\
Doe, Peter, University of Washington \\
Efremenko, Yuri, Oak Ridge National Laboratory \\
Ejiri, Hiroyasu, Osaka University \\
Elliott, Steve, University of Washington \\
Fardon, Rob, University of Washington \\
Fazely, Ali, Southern University \\
Freedman, Stuart J., Lawrence Berkeley National Laboratory \\
Fuller, George M., UC San Diego \\
Garcia, Alejandro, University of Notre Dame \\
Garvey, Gerry, Los Alamos National Laboratory \\
Goldman, Terrance J., Los Alamos National Laboratory \\
Haines, Todd, Los Alamos National Laboratory \\
Hamer, Andre, Los Alamos National Laboratory \\
Hartmann, Frank, Virginia Tech/NRL \\
Haxton, Wick, University of Washington \\
Hazama, Ryuta, University of Washington \\
Heeger, Karsten, University of Washington \\
Heise, Jaret, University of British Columbia \\
Hix, William, Oak Ridge National Laboratory \\
Hornish, Michael, Duke University \\
Horowitz, Charles, Indiana University \\
Ji, Xiangdong, University of Maryland \\
Jung, Chang Kee, SUNY Stony Brook \\
Keister, Bradley, National Science Foundation \\
Kolbe, Edwin, Institute for Theoretical Physics, Basel \\
Kouzes, Richard, Pacific Northwest National Laboratory \\
Krastev, Plamen, Institute for Advanced Study \\
Kusenko, Alexander, UCLA \\
Lande, Kenneth, University of Pennsylvania \\
Lanou, Jr., Robert, Brown University \\
Lesko, Kevin, Lawrence Berkeley National Laboratory \\
Loh, Eugene, National Science Foundation \\
Louis, William C., Los Alamos National Laboratory \\
Matzner, Richard A., University of Texas \\
McDonald, Art, Queen's University \\
McKeown, Robert, Caltech \\
McKinsey, Dan, Harvard University \\
McLaughlin, Gail, SUNY Stony Brook \\
Meyer, Bradley, Clemson University \\
Mezzacappa, Anthony, Oak Ridge National Laboratory \\
Miley, Harry, Pacific Northwest National Laboratory \\
Mills, Geoffrey B., Los Alamos National Laboratory \\
Murphy, Alex, Ohio State University \\
Nelson, Donald J. , University of Washington \\
Nelson, Roger, Department of Energy \\
Ng, John, TRIUMF/UBC \\
Norsen, Travis, University of Washington \\
Okada, Colin, Lawrence Berkeley National Laboratory \\
Orrell, John, University of Washington \\
Parker, Peter, Yale University \\
Poon, Alan, Lawrence Berkeley National Laboratory \\
Qian, Yong-Zhong, University of Minnesota \\
Raghavan, Raju S., Bell Laboratories \\
Reddy, Sanjay, University of Washington \\
Robertson, Hamish, University of Washington \\
Savage, Martin, University of Washington \\
Seki, Ryoichi, Cal State Northridge \\
Shutt, Tom, Princeton University \\
Snover, Kurt, University of Washington \\
Stefanski, Ray, FermiLab \\
Stoeffl, Wolfgang, Lawrence Livermore National Laboratory \\
Stokstad, Robert G., Lawrence Berkeley National Laboratory \\
Stonehill, Laura, University of Washington \\
Strayer , Michael, Oak Ridge National Laboratory \\
Sutton, C. Sean, Mount Holyoke College \\
Tayloe, Rex, Los Alamos National Laboratory \\
Thomas, Jeremy, University of Washington \\
Tornow, Werner, Duke University \\
Towe, Joseph, Antelope Valley College \\
Vagins, Mark, UC Irvine \\
Vernon, Wayne, UC San Diego \\
Vogel, Petr, Caltech \\
Vogelaar, R Bruce, Virginia Tech \\
Weinheimer, Christian, Johannes Gutenberg University \\
Wildenhain, Paul, University of Pennsylvania \\
Wilkerson, John, University of Washington \\
Wilson, James R., Lawrence Livermore National Laboratory \\
Zafonte, Steven, University of Washington \\
\end{center}

\end{document}